\documentclass{aa}
\usepackage[varg]{txfonts}
\usepackage{graphicx}
\usepackage{natbib}
\usepackage{listings}
\usepackage{comment}
\usepackage{multirow}
\usepackage{color}
\bibpunct{(}{)}{;}{a}{}{,}

\begin{document}

\title{Comparison between the first and second mass eruptions from progenitors of Type IIn supernovae}
\author{Naoto Kuriyama\inst{\ref{inst1},\ref{inst2}}
\and Toshikazu Shigeyama\inst{\ref{inst1},\ref{inst2}}}
\institute{Research Center for the Early Universe, Graduate School of Science, University of Tokyo, Bunkyo-ku, Tokyo, Japan\label{inst1}
\and
Department of Astronomy, Graduate School of Science, University of Tokyo, Bunkyo-ku, Tokyo, Japan\label{inst2}
}
\date{Received 11 June 2020 / Accepted 16 December 2020}
\abstract {Some massive stars experience episodic and intense mass loss phases with fluctuations in the luminosity. Ejected material forms circumstellar matter around the star, and the subsequent core collapse results in a Type IIn supernova that is characterized by interaction between supernova ejecta and circumstellar matter. The energy source that triggers these mass eruptions and dynamics of the outflow have not been clearly explained. Moreover, the mass eruption itself can alter the density structure of the envelope and affect the dynamics of the subsequent mass eruption if these events are repeated. A large amount of observational evidence suggests multiple mass eruptions prior to core collapse.}{We investigate the density structure of the envelope altered by the first mass eruption and the nature of the subsequent second mass eruption event in comparison with the first event.}{We deposited extra energy at the bottom of the hydrogen envelope of 15$M_\odot$ stars twice and calculated the time evolution by radiation hydrodynamical simulation code. We did not deal with the origin of the energy source, but focused on the dynamics of repeated mass eruptions from a single massive star.}{There are significant differences between the first and second mass eruptions in terms of the luminosity, color, and amount of produced circumstellar matter. The second eruption leads to a redder burst event in which the associated brightening phase lasts longer than the first. The amount of ejected matter is different even with the same deposited energy in the first and second event, but the difference depends on the density structure of the star.}{Upcoming high cadence and deep transient surveys will provide us a lot of pre-supernova activities, and some of which might show multi-peaked light curves. These should be interpreted taking the effect of density structure altered by the preceding outburst events into consideration.}
\titlerunning{The first and second mass eruptions}
\authorrunning{Kuriyama and Shigeyama}
\maketitle

\section{Introduction}
Growing observational evidence suggests that massive stars sometimes experience episodic and intense mass loss accompanied by temporal brightening. Eta Carinae is one of the most well-studied and well-known objects that experienced such an event \citep[e.g., ][]{1997ARA&A..35....1D}. This intense mass loss or brightening event has been considered to be related with the activity of luminous blue variables (LBVs), which were introduced by \citet{1984IAUS..105..233C}. On the other hand, some recent observations suggest that Wolf-Rayet stars (WR stars) may also experience such events \citep{2007ApJ...657L.105F, 2008MNRAS.389..131P, 2020MNRAS.492.5897S}.

These mass loss events could affect the observational features of subsequent core-collapse supernovae (SN). After a massive star abruptly loses a significant portion of the envelope, dense circumstellar matter (CSM) is formed around the star. If a core-collapse SN (CCSN) takes place in this dense CSM, the ejecta collide with the CSM. Then the kinetic energy of the ejecta is dissipated at shocks and becomes the main energy source \citep[e.g.,][]{1992SvA....36...63C, 2017hsn..book..403S}. These kinds of SNe are classified as Type IIn supernovae (SNe IIn) in case of hydrogen-rich CSM \citep{1990MNRAS.244..269S} or Type Ibn supernovae (SNe Ibn) in case of helium-rich CSM \citep{2007Natur.447..829P}.

Recent observations have revealed that some progenitors of SNe IIn experienced temporary brightening phase a few years or a few decades before the core collapse. For example, SN 2018cnf \citep{2019A&A...628A..93P}, SN 2016bdu \citep{2018MNRAS.474..197P}, SN 2013gc \citep{2019MNRAS.482.2750R}, PTF12cxj \citep{2014ApJ...789..104O}, and SN 2011ht \citep{2013ApJ...779L...8F} were reported to exhibit such brightening. Since the peak luminosity exceeds the Eddington luminosity of a massive star, this brightening must lead to an eruptive mass loss. The sparsely observed light curves prior to these SNe show that the brightening phase typically lasted for several years and indicate that the progenitors repeated episodic mass loss events during this period. SN 2009ip is one of the most famous SNe IIn whose progenitor star experienced multiple brightening phases likely associated with episodic mass loss events. The progenitor star of SN 2009ip in a brightening phase was first detected in 2009, and repeatedly exhibited brightening with short intervals less than 50 days in 2011 \citep{2013ApJ...767....1P}. Eventually this object experienced the most luminous outburst "2012b" in 2012. While some authors have argued that it was a genuine CCSN \citep[e.g.,][]{2013MNRAS.430.1801M, 2014MNRAS.438.1191S, 2017MNRAS.469.1559G}, some have suggested other scenarios (e.g., merger-burst event \citep{2013ApJ...764L...6S}, pulsational pair-instability event \citep{2013ApJ...767....1P}). The progenitor mass of SN 2009ip is estimated as $50-80\,M_\odot$ \citep{2010AJ....139.1451S}.

The mass loss rates from progenitors of SNe IIn are estimated at $0.026-0.12M_\odot\ \mathrm{yr}^{-1}$ \citep{2012ApJ...744...10K}, $10^{-4}-10^{-2}M_\odot\ \mathrm{yr}^{-1}$ \citep{2013A&A...555A..10T}, or more than $10^{-3}M_\odot$ \citep{2014MNRAS.439.2917M}. These values are so high that they cannot be reconciled with a steady wind mass loss model \citep{2001A&A...369..574V, 2005A&A...438..273V, 2014ARA&A..52..487S}. In addition, some SNe IIn show bumps in their light curves that are thought to be related to bumpy density structures of CSM \citep{2019MNRAS.482.2750R, 2017A&A...605A...6N, 2012ApJ...756..173S}. Although these bumps seem to be rare cases among SNe IIn \citep{2019arXiv190605812N}, this bumpy CSM structure also implies not steady but episodic mass loss events from the progenitor star. Therefore, a dynamical phenomenon, which is not included in most of current stellar evolution models, would occur during the temporary brightening phase.

However, the extra energy sources that trigger these mass loss events and dynamical mechanism of the mass eruption reproducing observational features have not been fully understood. There are several works that investigate the dynamics of the stellar envelope of massive stars into which an extra energy is deposited \citep{2016MNRAS.458.1214Q,2017MNRAS.470.1642F,2018MNRAS.476.1853F,2019ApJ...877...92O,2010MNRAS.405.2113D,2019MNRAS.485..988O,2020A&A...635A.127K}. In \citet{2020A&A...635A.127K}, we investigated the observational properties of mass eruption from the progenitors of SNe IIn/Ibn assuming an extra energy is supplied into the envelope of single massive stars. We deposited energy into the bottom of the stellar envelope with a short timescale and calculated the dynamics of the subsequent mass eruption using 1-D radiation hydrodynamical simulation code.

Although energy is deposited only once and the corresponding single dynamical eruption event is discussed in these studies including \citet{2020A&A...635A.127K}, an intense mass loss event is often repeated in real situations, as discussed above, and this could become a key factor for understanding the mass loss mechanism. Once a dynamical eruption takes place, the expanding envelope keeps its altered density profile for the thermal timescale. If another eruption event occurs again before the envelope returns to a hydrostatic equilibrium state, the property of the second eruption may be completely different because of a different density profile of the envelope.

\begin{figure}
\resizebox{\hsize}{!}{\includegraphics{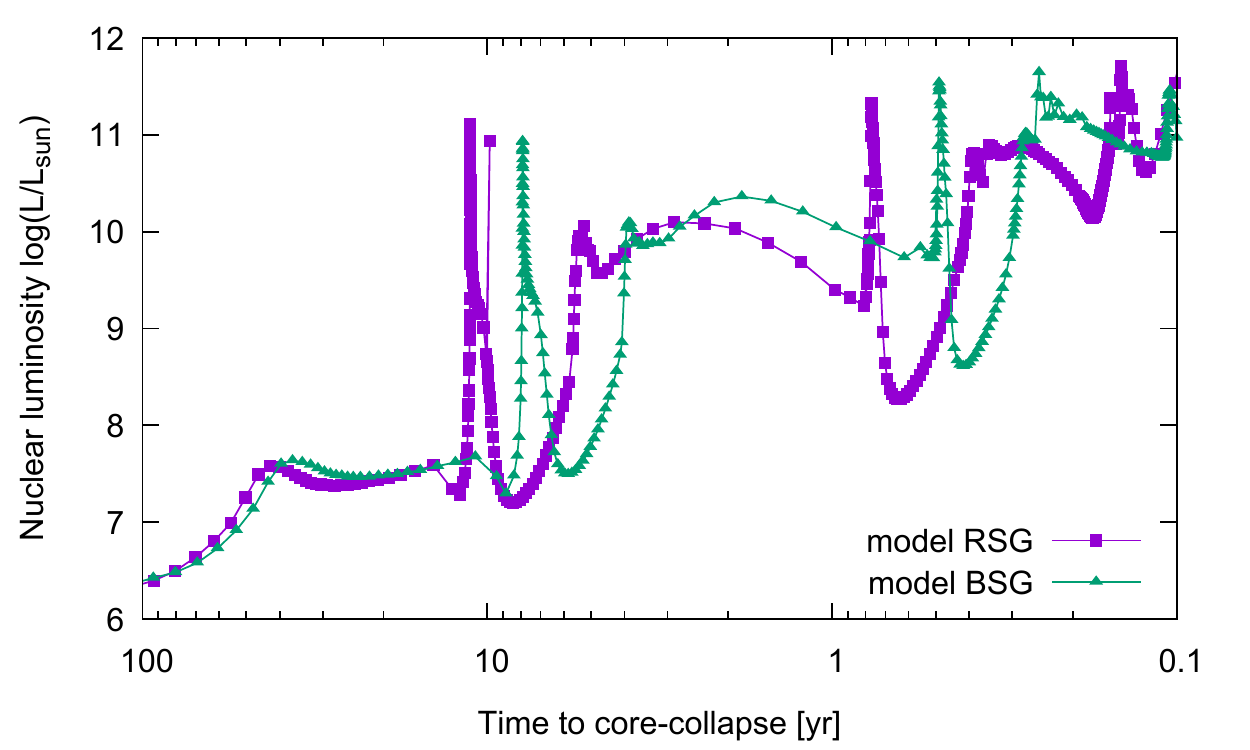}}
\caption{Evolution of the energy generation rate by nuclear burning for each model indicated by labels. The rate becomes higher toward the core collapse. There are some local peaks for each model around ten years, five years, and one year before core collapse. We adopted the peaks at 11.2 years before core collapse (model RSG) and at 7.2 years before core collapse (BSG) as initial models for our calculations. These peaks correspond to the core neon burning phase.}
\label{NuclearLuminosity}
\end{figure}

To deal with this problem, in this work we study the dynamics of eruptive mass loss that is repeated twice, as a supplemental work of \citet{2020A&A...635A.127K}. Of course eruption can repeat more than twice, but we just focus on a comparison between eruptions from the original (hydrostatic) envelope and the expanding envelope. We assume that eruptive mass loss is related to the local peak of the energy generation rate of nuclear burning shown in Figure \ref{NuclearLuminosity}, although the specific mechanism of energy transportation from the burning region to the envelope is not assumed as was the case in \citet{2020A&A...635A.127K}. We deposit extra energy at the bottom of the hydrogen envelope twice and investigate the properties of each eruption and difference between the first and second eruptions. The second energy injection is conducted before the stellar envelope relaxes from the first injection. Each extra energy is injected for a short period of time and the resulting dynamical eruption is investigated by radiation hydrodynamical simulations. In Sect. 2, we introduce the progenitor models, which we use as initial models in our simulation and the method of radiation hydrodynamical calculation. In Sect.3 we present the results of calculation and find out the differences between the first and second eruptions. We clarify the implications to the observations in Sect. 4 and present our conclusions in Sect. 5.

\begin{table*}
\caption{Properties of two SNe Progenitor models.}
\label{table:1}
\centering
\begin{tabular}{c c c c c c c c c c c}
\hline\hline
Model & $M_\mathrm{ZAMS}$ & $Z$ & $R$ & $T_\mathrm{eff}$ & $M_\mathrm{He\ core}$ & $M_\mathrm{H\ env}$ & $\Gamma_\mathrm{e}$\tablefootmark{a} & $E_\mathrm{envelope}$\tablefootmark{b} & Time to CC & Burning stage\\
\hline
RSG & $15M_\odot$ & $0.02$ & $696R_\odot$ & $3500$K & $4.1M_\odot$ & $10.6M_\odot$ & 0.07 & $-5.6\times 10^{47}$ erg& $11.2$ yr & Ne burning\\
BSG & $15M_\odot$ & $0.0002$ & $76R_\odot$ & $10200$K & $4.2M_\odot$ & $10.8M_\odot$ & 0.36 & $-1.7\times 10^{49}$ erg& $7.9$ yr & Ne burning\\
\hline
\end{tabular}
\tablefoot{
\tablefootmark{a}{Eddington factor considering electron scattering opacity.}
\tablefootmark{b}{Total energy of H-rich envelope.}
}
\end{table*}

\begin{table*}
\caption{Amount of injected energy $E_\mathrm{inj1}$, $E_\mathrm{inj2}$; $E_\mathrm{inj1}$ at $t = 0$ and $E_\mathrm{inj2}$ at $t = \Delta t_\mathrm{inj}$ are injected. Five patterns of calculations were conducted for each model (a total of 10 patterns of calculations).}
\label{table:2}
\centering
\begin{tabular}{c c c c c}
\hline\hline
Calculation model & Progenitor model & $E_\mathrm{inj1}$ [erg]& $E_\mathrm{inj2}$ [erg] & $\Delta t_\mathrm{inj}$ \\
\hline
RSG1-f (fiducial) & \multirow{5}{*}{RSG} & $1.5\times 10^{47}$ & $1.5\times 10^{47}$ &$1.0t_\mathrm{dyn}$ (98 days) \\
RSG1-s (short) & & $1.5\times 10^{47}$ & $1.5\times 10^{47}$ &$0.5t_\mathrm{dyn}$ (49 days)\\ 
RSG1-m (medium) & & $1.5\times 10^{47}$ & $1.5\times 10^{47}$ &$2.0t_\mathrm{dyn}$ (196 days)\\ 
RSG1-l (long) & & $1.5\times 10^{47}$ & $1.5\times 10^{47}$ &$4.0t_\mathrm{dyn}$ (392 days)\\ 
RSG2-f (fiducial) & & $1.5\times 10^{47}$ & $3.0\times 10^{47}$ &$1.0t_\mathrm{dyn}$ (98 days)\\ \hline
BSG1-f (fiducial) & \multirow{5}{*}{BSG} & $6.0\times 10^{48}$ & $6.0\times 10^{48}$ &$1.0t_\mathrm{dyn}$ ($3.5$ days)\\
BSG1-s (short) & & $6.0\times 10^{48}$ & $6.0\times 10^{48}$ &$0.5t_\mathrm{dyn}$ ($1.8$ days)\\
BSG1-m (medium) & & $6.0\times 10^{48}$ & $6.0\times 10^{48}$ &$2.0t_\mathrm{dyn}$ ($7.1$ days)\\
BSG1-l (long) & & $6.0\times 10^{48}$ & $6.0\times 10^{48}$ &$4.0t_\mathrm{dyn}$ ($14$ days)\\
BSG2-f (fiducial) & & $6.0\times 10^{48}$ & $9.0\times 10^{48}$ &$1.0t_\mathrm{dyn}$ ($3.5$ days)\\
\hline
\end{tabular}
\end{table*}

\begin{figure}
\resizebox{\hsize}{!}{\includegraphics{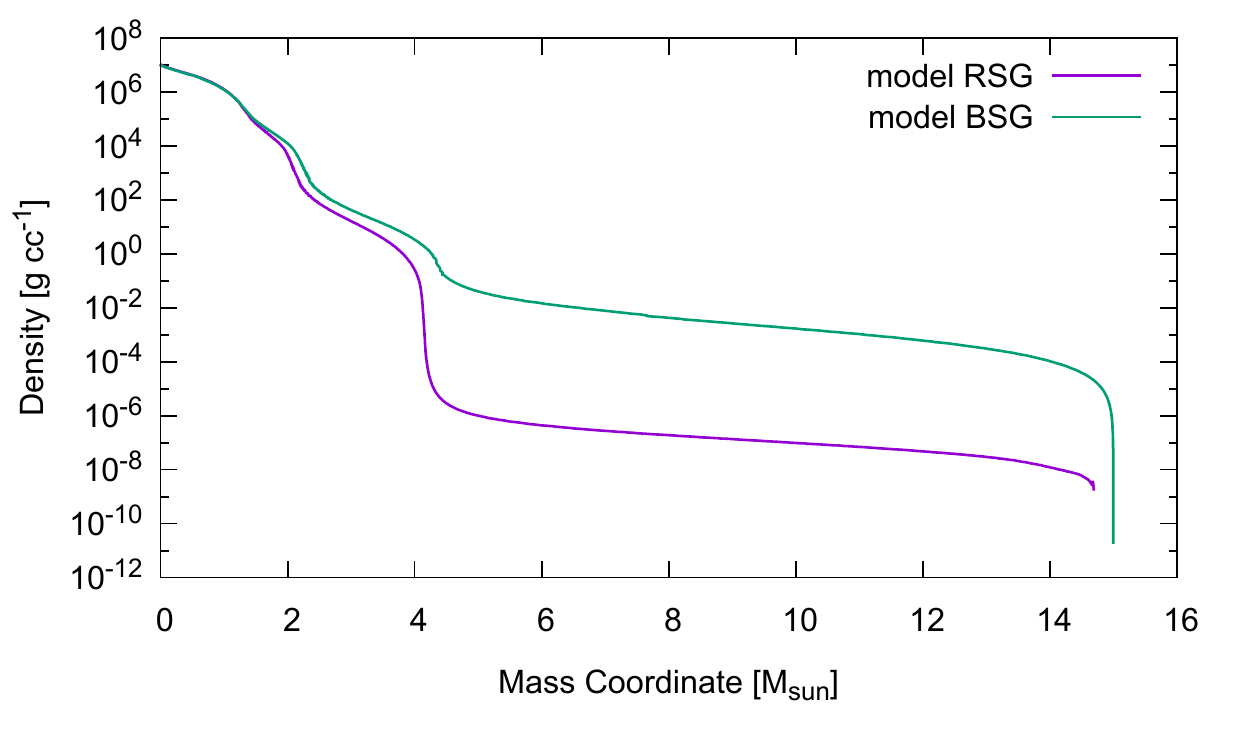}}
\caption{Density distribution of each progenitor model. Model RSG has a more extended envelope than model BSG because of higher metallicity (Table \ref{table:1}) and opacity. In addition to that, model RSG has experienced stronger mass loss due to higher metallicity and thus has a lower total mass.}
\label{DensityProfile}
\end{figure}

\section{Setup and methods}
\subsection{Progenitor models}
Although it has been often considered that eruptive mass loss and SNe IIn are related to activities of LBVs \citep[e.g., ][]{2006A&A...460L...5K, 2007ApJ...656..372G, 2012ARA&A..50..107L} as described in Sect 1, observations suggest that red supergiants (RSGs) may also be progenitors of SNe IIn \citep{2009AJ....137.3558S, 2015MNRAS.450..246B}. The progenitor mass is also ambiguous and varies with the subclass of SNe IIn \citep{2014ARA&A..52..487S}. In this work, we adopted two types of $15M_\odot$ progenitors, namely the blue supergiant model (model BSG) and the red supergiant model (model RSG). We made these progenitor models using the stellar evolution code MESA (release 10398) \citep{2011ApJS..192....3P, 2013ApJS..208....4P, 2015ApJS..220...15P, 2018ApJS..234...34P, 2019ApJS..243...10P}. We did not include the effect of rotation in our calculations. We stopped the calculations of MESA 11.2 yr before core collapse for model RSG and 7.9 yr before core collapse for model BSG, respectively (Fig. \ref{NuclearLuminosity}). We remeshed these models to resolve the structure of radiation mediated shocks in the outer envelope and used them as initial models for our hydrodynamical simulations presented in Sect. 2.2. We set the numbers of grid points to 10000 cells for model RSG and 5000 cells for model BSG. We tested other mass resolutions for model RSG (8000, 12000, 14000, and 16000 cells), and we find that as the number of grid points becomes larger the amount of ejected mass by energy injection tends to increase. The increase converges above 14000 cells, and there is a roughly ten percent difference in ejected mass between in case of 10000 cells (the resolution adopted in this paper) and 16000 cells. Therefore, model RSG has an ejected mass accuracy of ten percent at worst. Model BSG has more cells in the ejected matter than model RSG. Thus we think model BSG has an ejected mass accuracy comparable to or better than model RSG.

These two models have the same zero-age main sequence mass ($15\,M_\odot$) but different metallicities (Table \ref{table:1}). The difference in metallicity causes different opacities in the stellar envelope. As a consequence, a star with higher metallicity evolves to a RSG and another star with lower metallicity becomes a BSG (different density distributions in Fig. \ref{DensityProfile}; radii and effective temperatures in Table \ref{table:1}). The time evolution of the energy generation rate of nuclear burning is also slightly different between the two models. The timescale of each nuclear burning stage in model RSG is longer than that in model BSG (Fig. \ref{NuclearLuminosity}) because model RSG has a lower core mass (Table \ref{table:1}). Model RSG has experienced more intense steady mass loss because of higher metallicity and thus has a smaller core. We started hydrodynamical calculations at 11.2 yr before core collapse for models RSG and 7.9 yr for models BSG corresponding to the local peaks of nuclear burning luminosity, as mentioned in \S 2.2. Although we can identify other higher peaks within one year before core collapse, we only focus on these peaks because it could take more than one year for the ejected envelope to extending far enough to reproduce the observed typical distance from the progenitor star to a CSM \citep[e.g., 160 AU,][]{2007ApJ...671L..17S}.
Detailed methods of constructing the two progenitor models are described in Appendix A.

\subsection{1-D radiation hydrodynamical simulation}
As described in Sect. 1, the main purpose of this paper is to reveal differences in the properties between the first and second mass eruptions from SNe IIn progenitors. To investigate this topic, we used the same 1-D radiation hydrodynamical simulation method as in \citet{2020A&A...635A.127K}. This code was originally developed by \citet{1990ApJ...360..242S} and was revised to solve the Riemann problem by the piecewise parabolic method \citep{1984JCoPh..54..174C}; radiation transfer is treated by flux-limited diffusion approximation \citep{1981ApJ...248..321L}. We used Kramer's opacity and did not include the effect of dust formation. Therefore, a calculation of a low temperature region in which the dust opacity becomes important should contain some uncertainty. We injected extra energy into the bottom of the hydrogen envelope twice without specifying from where and how the extra energy is supplied; the mechanism for supplying the extra energy has not been fully understood so far (see Sect. 1). The interval between the first and the second energy injection $\Delta t$ was scaled with the dynamical timescales of the envelope $t_\mathrm{dyn}$ (98 days for RSG and 3.5 days for BSG). This is because we focus on the dynamics of the envelope, although the interval depends on the physics of energy source in reality. The eruption of the envelope proceeds on the timescale of $t_\mathrm{dyn}$ and after that it shrinks on the timescale of $t_\mathrm{KH}$. In our model, $t_\mathrm{KH}$ ($\sim 200\ \mathrm{yr}$ for model RSG and BSG) is much longer than the remaining time to core collapse, and thus only $t_\mathrm{dyn}$ seems to play a role in determining the physics of the multiple eruption. Therefore we used $t_\mathrm{dyn}$ as a fiducial interval between the two injections.
As shown in Table \ref{table:2}, the first energy $E_\mathrm{inj1}$ was injected at the time $t = 0$ and the second energy $E_\mathrm{inj2}$ at $t = \Delta t_\mathrm{inj}$. For each of the progenitor models RSG and BSG, five patterns of calculation with different amounts of the injected energy and $\Delta t_\mathrm{inj}$ were conducted (ten patterns of calculations in total). We call these models RSG1-f,s,m,l, RSG2-f, BSG1-f,s,m,l, and BSG2-f, respectively (Table \ref{table:2}). The letter f stands for "fiducial" and the other letters s, m, and l stand for short, medium, and long intervals, respectively. The quantity $E_\mathrm{inj1}$ was roughly set to one-third or one-quarter of $|E_\mathrm{envelope}|$ ($E_\mathrm{envelope}$ represents the total energy of the envelope) for every model. This value is enough to expel $10^{-2}-10^{-1}M_\odot$ material \citep{2020A&A...635A.127K}, which is a typical amount of CSM in SNe IIn.

\section{Results}

\begin{figure}
\resizebox{\hsize}{!}{\includegraphics{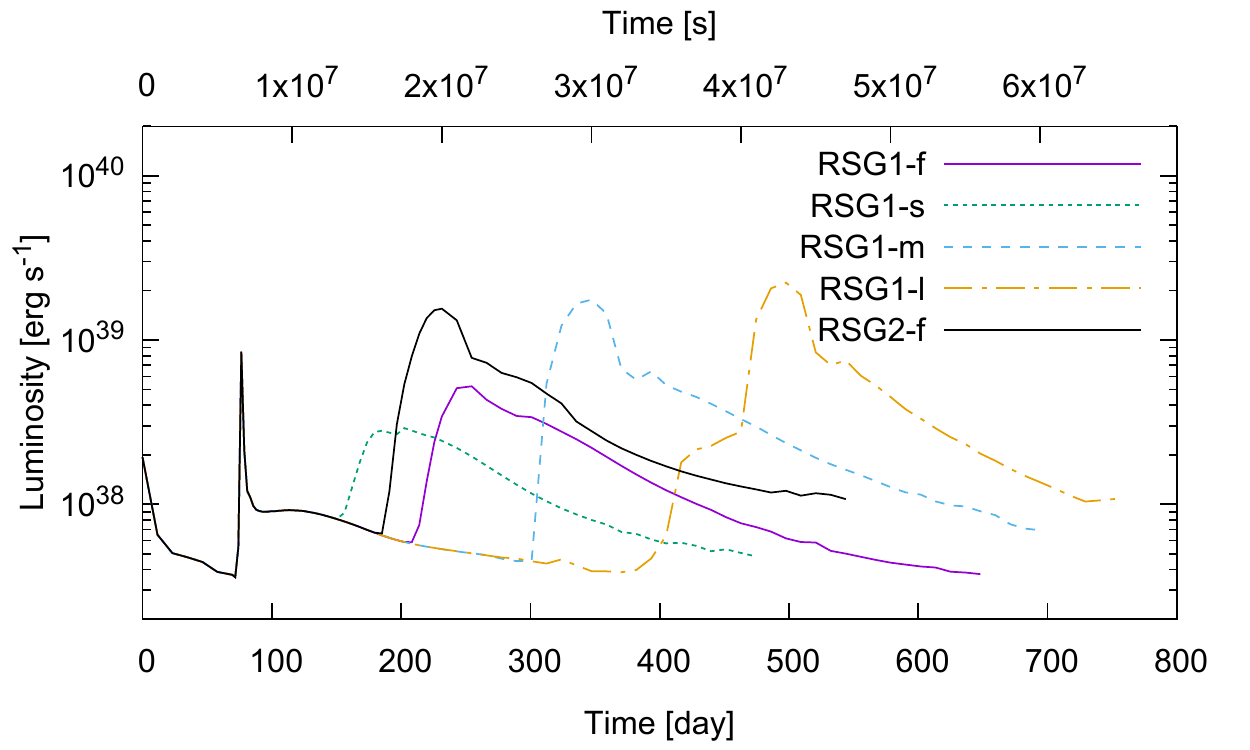}}
\caption{Light curves for models RSG1-f,s,m,l, and RSG2-f. The first eruption triggered by the injected energy $E_\mathrm{inj1}$ produces a peak at day $\sim$ 80. The second peak in each model corresponds to the second eruption triggered by the injected energy $E_\mathrm{inj2}$.}
\label{RSG_LC}
\end{figure}

\begin{figure}
\resizebox{\hsize}{!}{\includegraphics{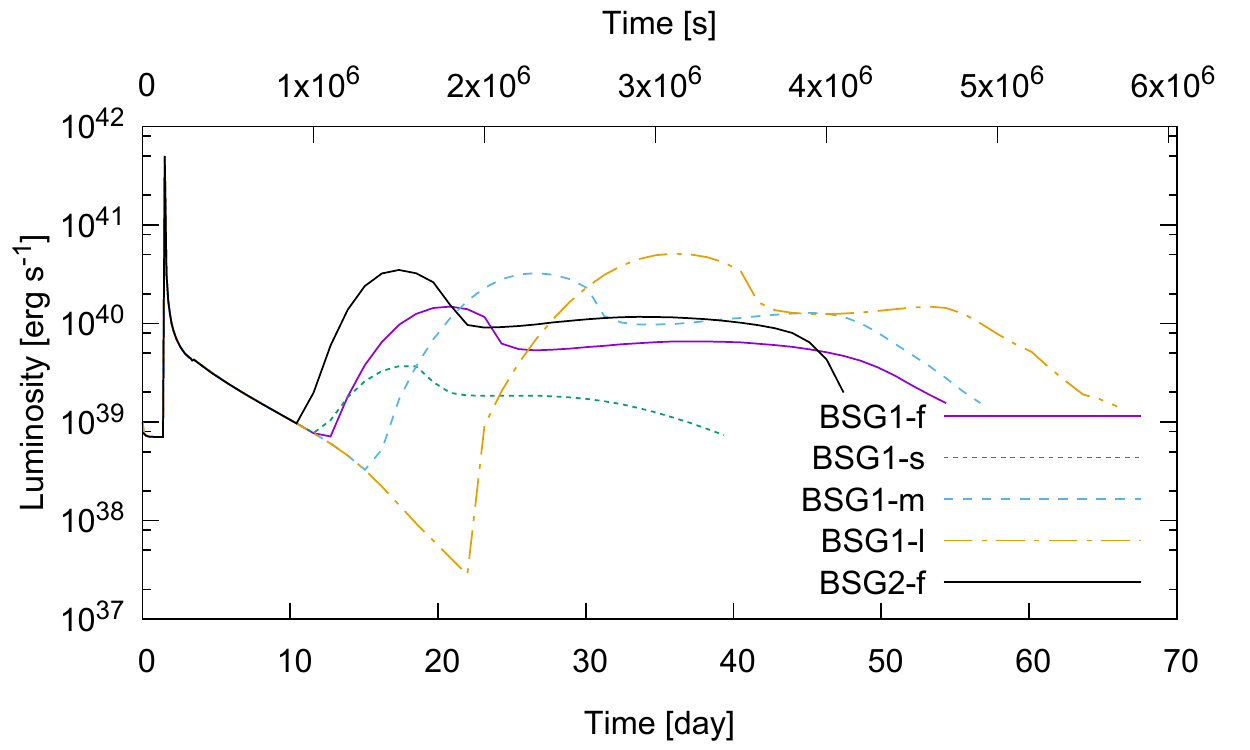}}
\caption{Same as Figure \ref{RSG_LC} but for models BSG1-f,s,m,l, and BSG2-f.}
\label{BSG_LC}
\end{figure}

\begin{figure}
\resizebox{\hsize}{!}{\includegraphics{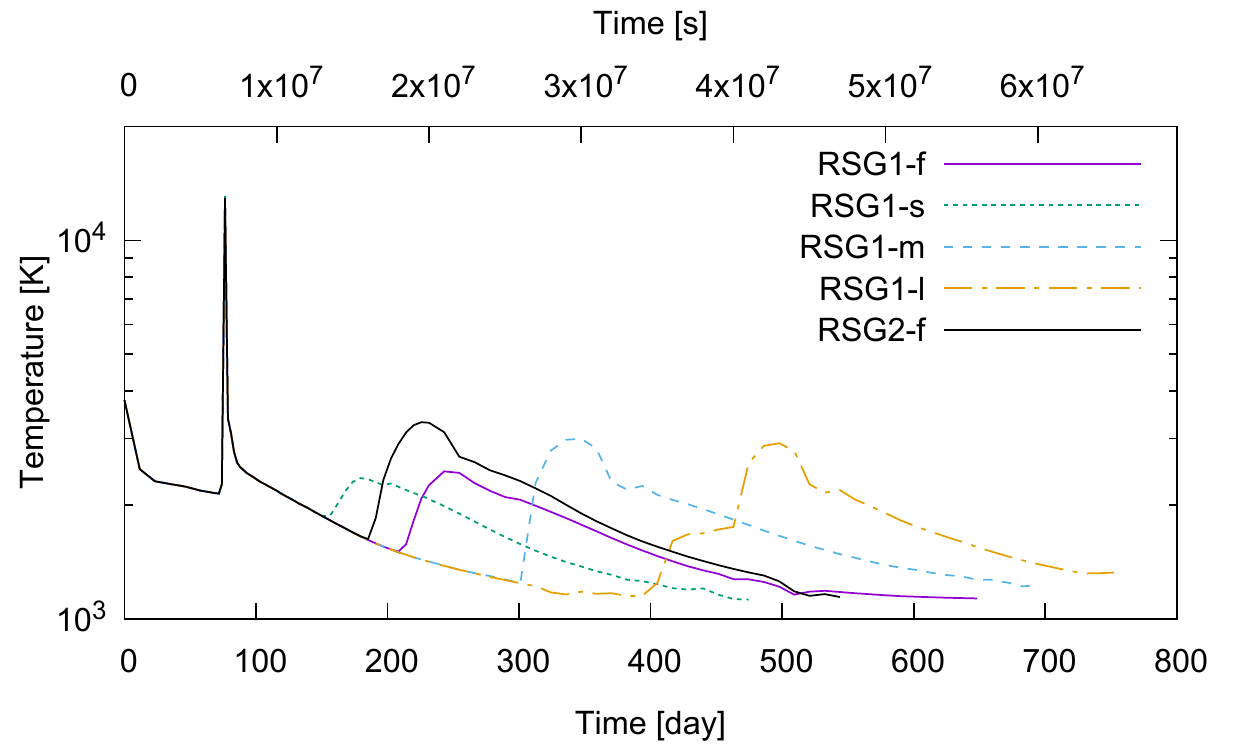}}
\caption{Evolution of effective temperature for each RSG progenitor model.}
\label{RSG_TEMP}
\end{figure}

\begin{figure}
\resizebox{\hsize}{!}{\includegraphics{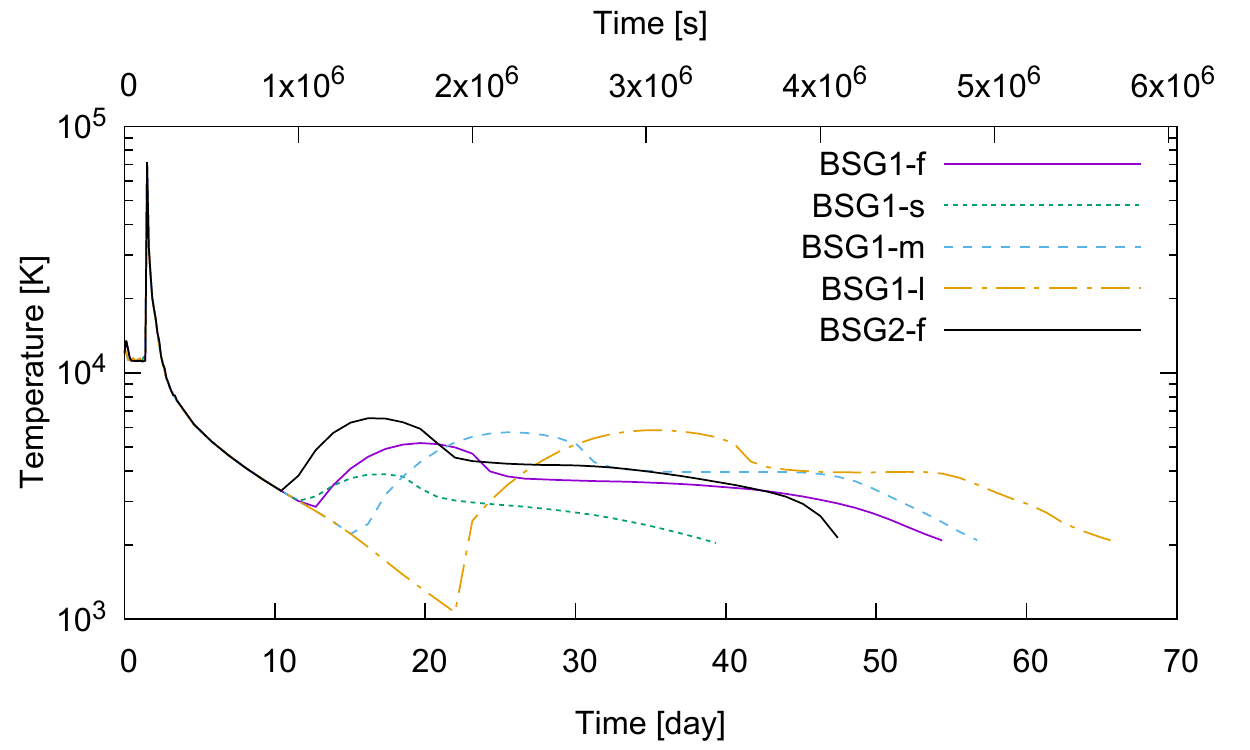}}
\caption{Evolution of effective temperature for each BSG progenitor model.}
\label{BSG_TEMP}
\end{figure}

\begin{table}
\caption{Amount of ejected mass and their kinetic energy for two distinct mass eruption events corresponding to the first and the second energy injection.}
\label{table:3}
\centering
\begin{tabular}{c c c}
\hline\hline
model & First eruption\tablefootmark{a} & Second eruption \\
\hline
RSG1-s & \multirow{5}{*}{$0.013 M_\odot$,$\ \ 2.8 \times 10^{44}\ \mathrm{erg}$} & $0.0011 M_\odot$,$\ \ 5.1 \times 10^{43}\ \mathrm{erg}$ \\
RSG1-f & & $0.0015 M_\odot$,$\ \ 4.7 \times 10^{43}\ \mathrm{erg}$ \\
RSG1-m & & $0.0013 M_\odot$,$\ \ 4.2 \times 10^{43}\ \mathrm{erg}$ \\
RSG1-l & & $0.0007 M_\odot$,$\ \ 3.0 \times 10^{43}\ \mathrm{erg}$ \\
RSG2-f & & $0.73 M_\odot$,$\ \ 1.5 \times 10^{46}\ \mathrm{erg}$ \\ \hline
BSG1-s & \multirow{5}{*}{$0.19 M_\odot$,$\ \ 2.2 \times 10^{47}\ \mathrm{erg}$} & $0.59 M_\odot$,$\ \ 2.5 \times 10^{47}\ \mathrm{erg}$ \\
BSG1-f & & $0.51 M_\odot$,$\ \ 3.4 \times 10^{47}\ \mathrm{erg}$ \\
BSG1-m & & $0.55 M_\odot$,$\ \ 3.3 \times 10^{47}\ \mathrm{erg}$ \\
BSG1-l & & $0.56 M_\odot$,$\ \ 3.5 \times 10^{47}\ \mathrm{erg}$ \\
BSG2-f & & $2.02 M_\odot$,$\ \ 8.4 \times 10^{48}\ \mathrm{erg}$ \\
\hline
\end{tabular}
\tablefoot{
\tablefootmark{a}{RSG1 and RSG2 (BSG1 and BSG2) have the same amount of ejected mass and kinetic energy in the first eruption because the same amount of energy $E_\mathrm{inj1}$ is injected.}
}
\end{table}

\begin{figure}
\resizebox{\hsize}{!}{\includegraphics{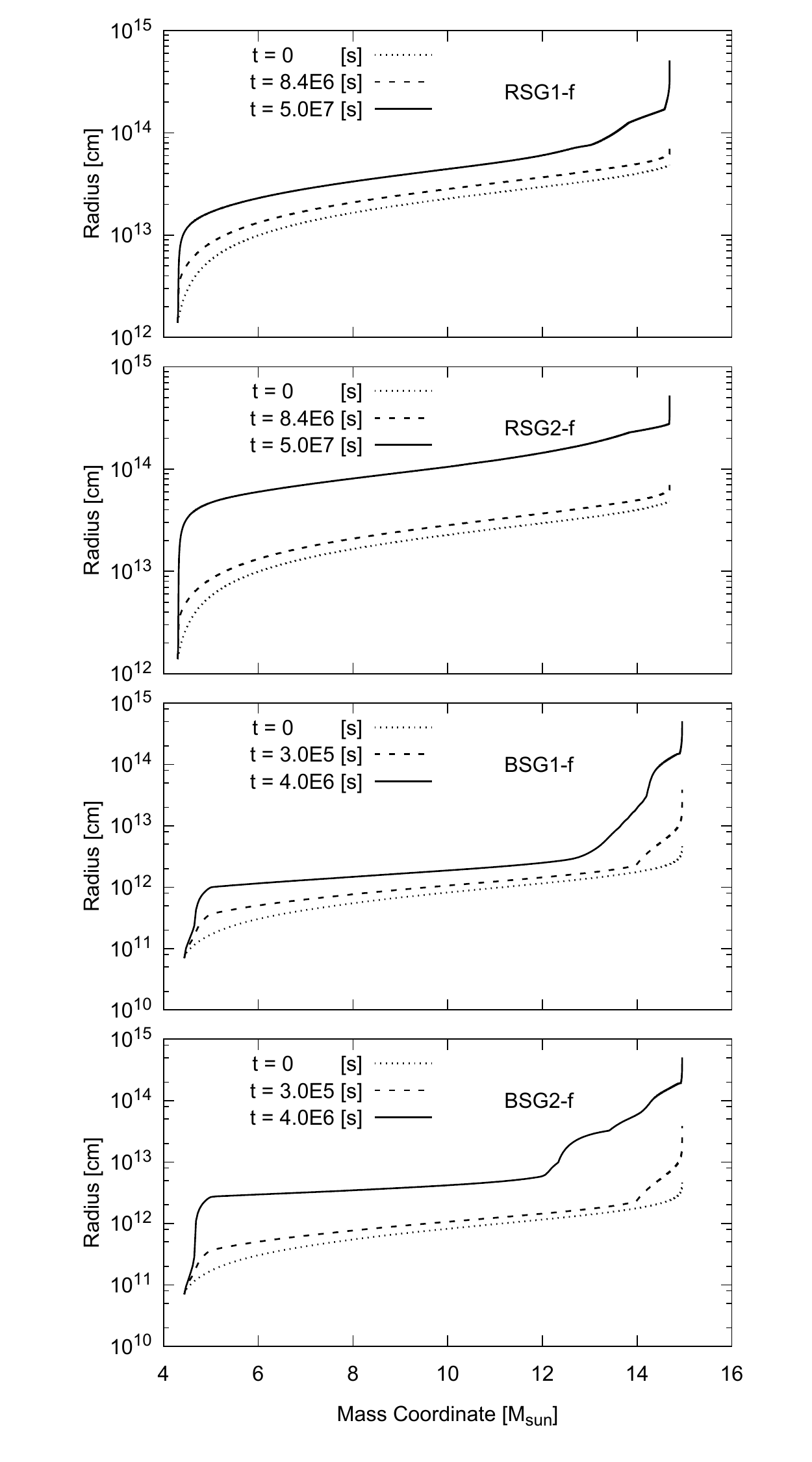}}
\caption{Radii as functions of the enclosed mass for the fiducial models. The three lines in each panel represent the profiles before the first energy injection (dotted line), between the first and the second energy injections (dashed line), and at the end of the hydrodynamical simulations (solid line).}
\label{Radius}
\end{figure}
\begin{figure}
\resizebox{\hsize}{!}{\includegraphics{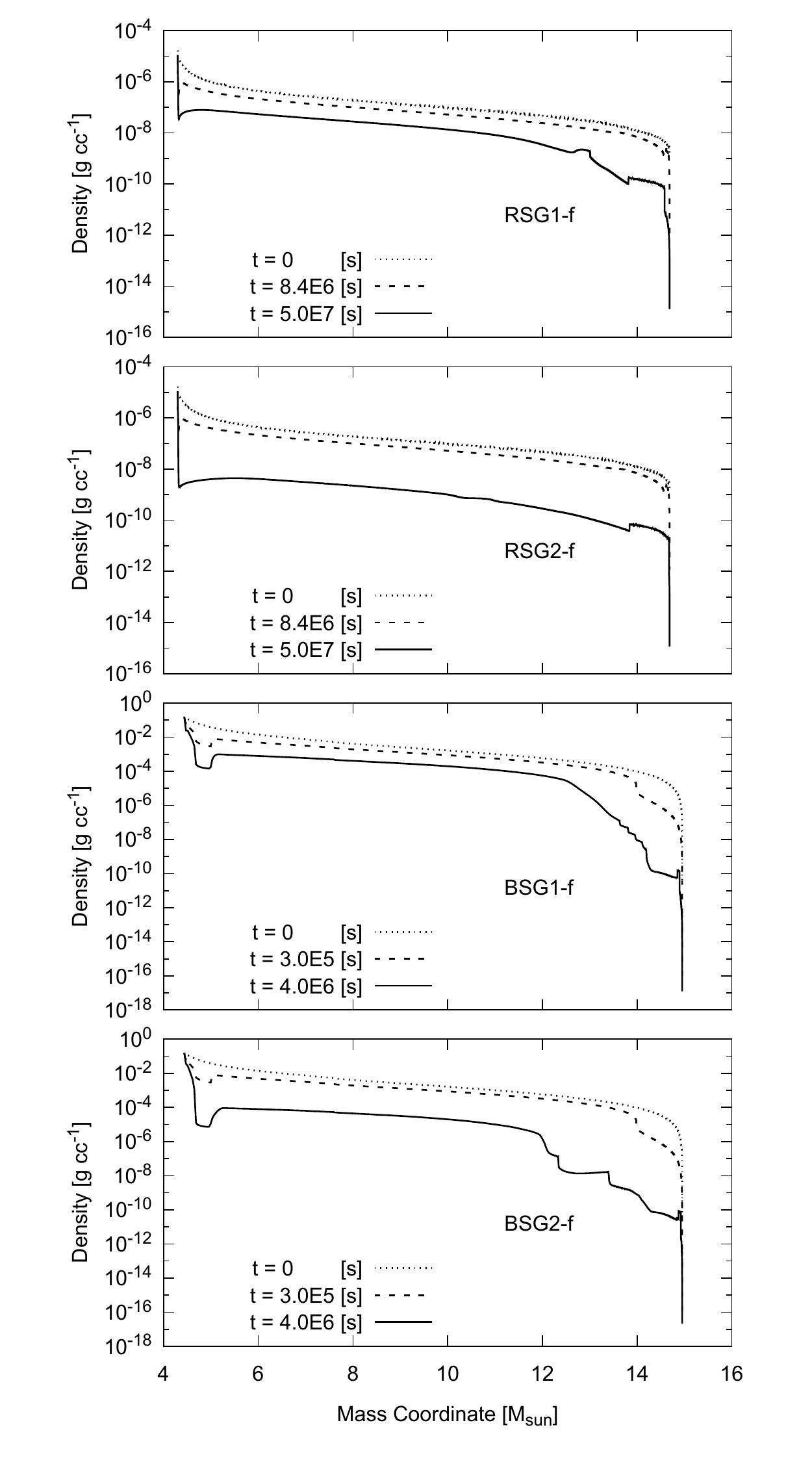}}
\caption{Same as figure \ref{Radius} but for the density profiles.}
\label{Density}
\end{figure}
\begin{figure}
\resizebox{\hsize}{!}{\includegraphics{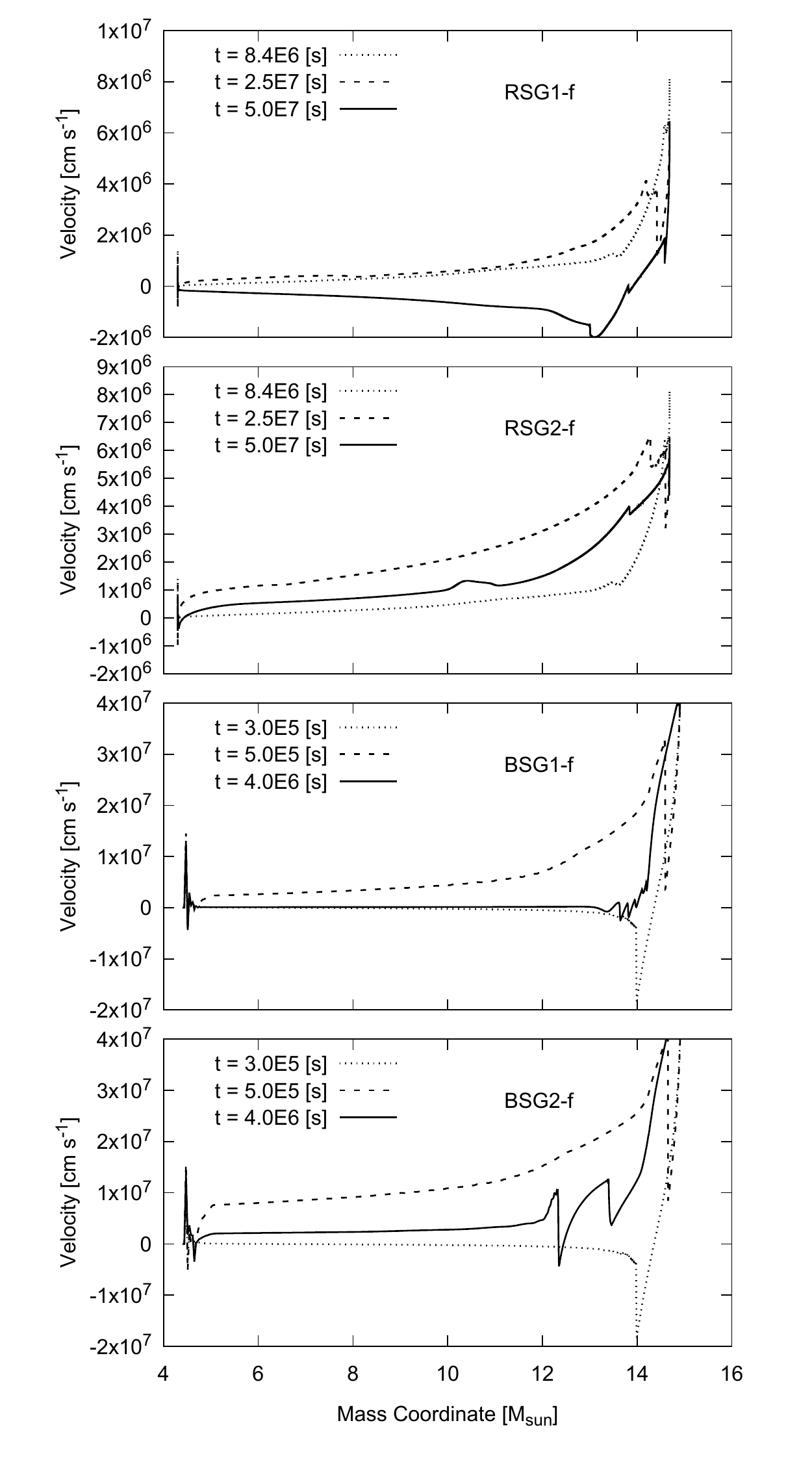}}
\caption{Velocity profiles as functions of the enclosed mass for the fiducial models. The three lines in each panel represent the profiles before the second energy injection (dotted line), after the second energy injection (dashed line), and at the end of the hydrodynamical simulations (solid line). }
\label{Velocity}
\end{figure}
\begin{figure}
\resizebox{\hsize}{!}{\includegraphics{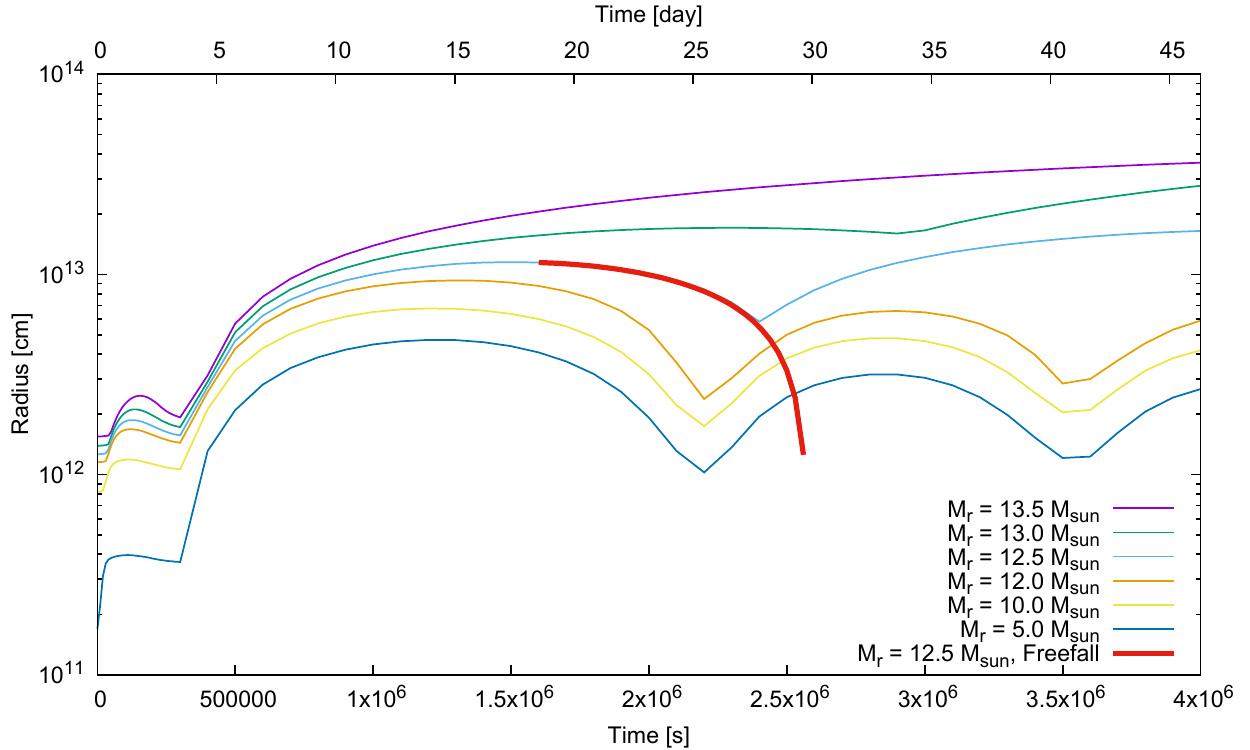}}
\caption{Trajectories of some fluid elements for model BSG2-f. The red bold line represents the trajectory of a matter with a mass coordinate $M_r = 12.5M_\odot$ assuming free-fall motion from the maximum radius.}
\label{trajectory}
\end{figure}
\begin{figure}
\resizebox{\hsize}{!}{\includegraphics{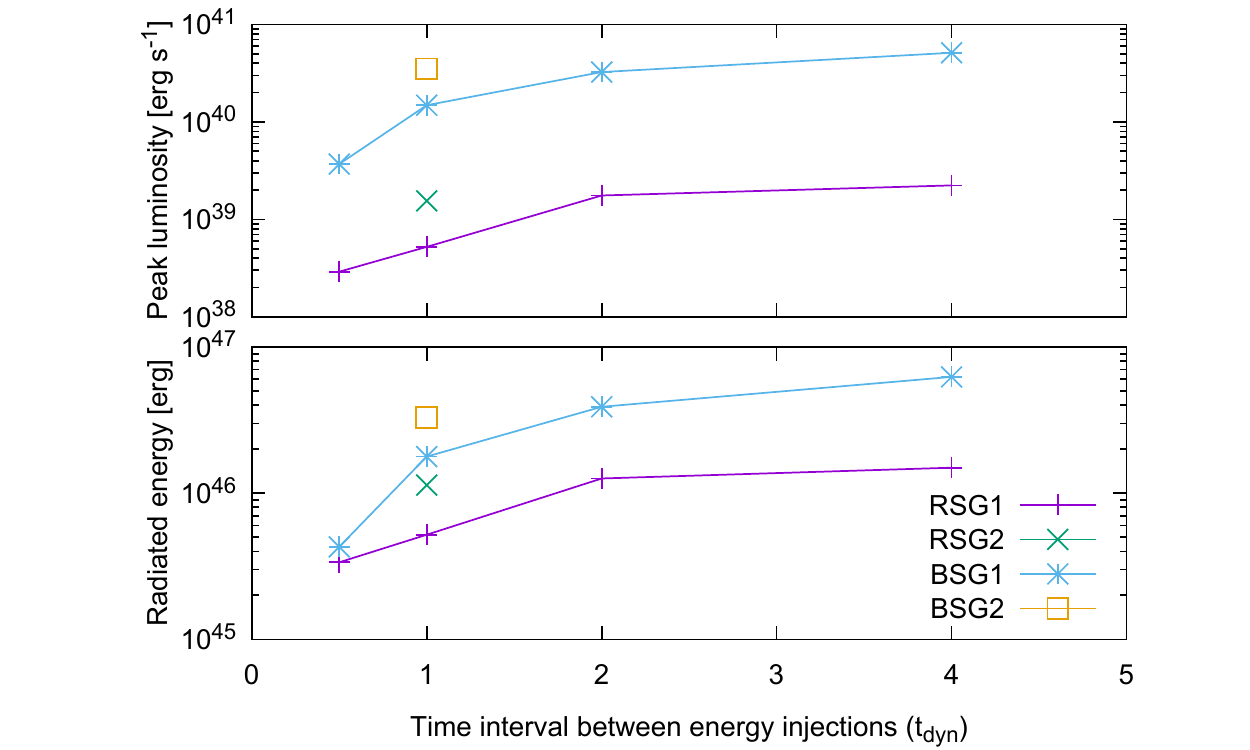}}
\caption{Peak luminosity (top panel) and total radiated energy (bottom panel) in the second eruption as functions of the interval $\Delta t_\mathrm{inj}$. The total radiated energy is evaluated by the cumulative luminosity from the time when the luminosity begins to rise to 200 days (20 days) after the peak for models RSG (BSG).}
\label{intervalCompare}
\end{figure}

\begin{figure}
\resizebox{\hsize}{!}{\includegraphics{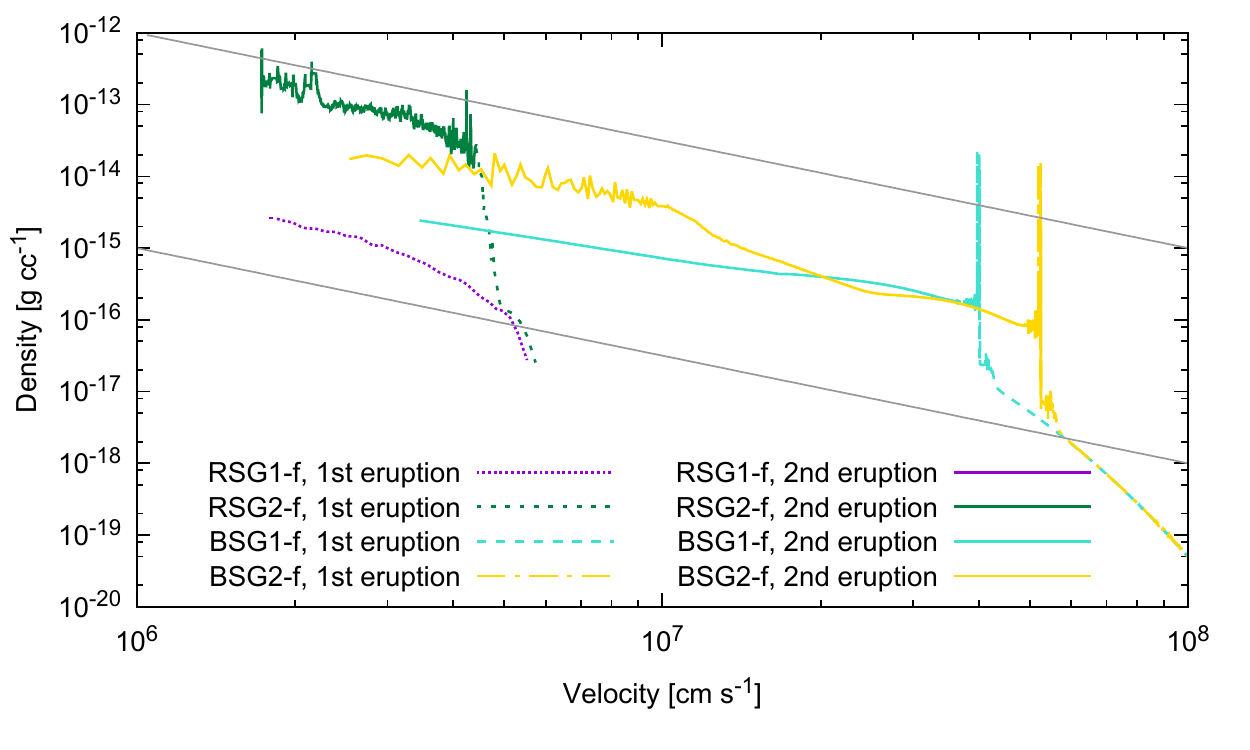}}
\caption{Density profiles of CSM as functions of velocity at the core collapse for the fiducial models. The broken lines (solid lines) correspond to the CSM erupted by the first injection (second injection). The gray line represents the relation $\rho \propto v^{-1.5}$, which we adopt as a fiducial slope in the case of eruptive mass loss \citep{2020A&A...635A.127K, 2020arXiv200506103T}.}
\label{DensityVelocity}
\end{figure}

\begin{figure*}
\resizebox{\hsize}{!}{\includegraphics{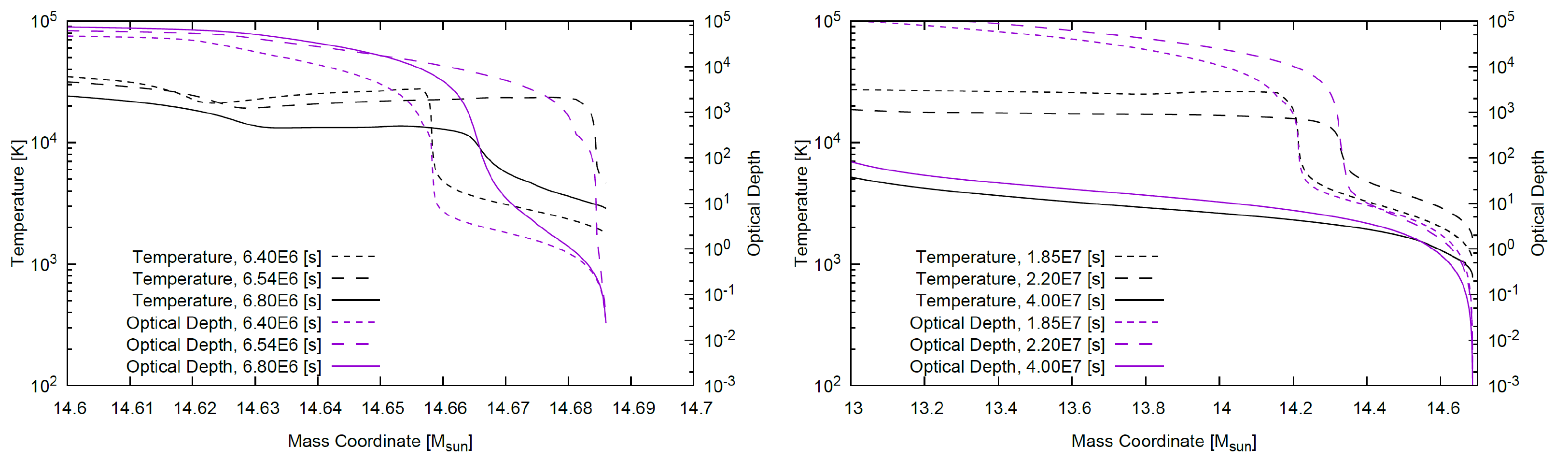}}
\caption{Temperature and optical depth for model RSG1-f during the first eruption (left) and the second eruption (right). Broken lines (6.4E6 in the left panel and 1.85E7 in the right panel) correspond to profiles just before eruptions, long-broken lines correspond to profiles at the time of peaks in the luminosity due to eruptions, and solid lines show the profiles after the luminosity peaks.}
\label{Eject_Detail}
\end{figure*}

\subsection{First energy injection and corresponding mass eruption}
Two injections of extra energies at different epochs cause two distinct mass eruption events (Table \ref{table:3}) and two distinct peaks of luminosity (Fig. \ref{RSG_LC}, \ref{BSG_LC}) and effective temperature (Fig. \ref{RSG_TEMP}, \ref{BSG_TEMP}). About one- quarter (model RSG1-f,s,m,l, and RSG2-f) or one-third (BSG1-f,s,m,l, and BSG2-f) of $|E_\mathrm{envelope}|$ was injected into the bottom of the hydrogen-rich envelope in the first energy injection $E_\mathrm{inj1}$. A relatively strong shock wave propagates toward the surface and breaks out with a rapidly rising light curve in all the models. The first energy injection ejects matter with masses of $\sim 0.01\,M_\odot$ (model RSG1-f,s,m,l and RSG2-f) and $\sim 0.2\,M_\odot$ (model BSG1-f,s,m,l and RSG2-f). The ejected mass in model RSG1-f,s,m,l and RSG2-f (model BSG1-f,s,m,l and BSG2-f) are equal in the first eruption because the same amount of energy $E_\mathrm{inj1}$ is injected. Although the ratio of $E_\mathrm{inj1}/E_\mathrm{envelope}$ for model RSG1-f and RSG2-f is similar to that for model BSG1-f and BSG2-f, there is an order of magnitude difference in the ejected mass above. This is because the  envelope of model BSG is denser than that of model RSG and therefore a more energetic shock wave propagates outward and expels a considerable amount of the envelope. For every model, the hydrogen-rich envelope remains inflated after the first eruption as indicated from the significantly altered radius and density profiles in Figures \ref{Radius} and \ref{Density}.

\subsection{Second energy injection and corresponding mass eruption}
Before the expanding envelope shrinks and returns to a hydrostatic equilibrium state, the second energy injection was conducted. The amount of injected energy is the same as in the first injection for models RSG1-f,s,m,l and BSG1-f,s,m,l. On the other hand, for models RSG2-f and BSG2-f, twice (model RSG2-f) or 1.5 times (model BSG2-f) more energy is injected. These second energy injections trigger the second mass eruption events accompanied with the second peaks in luminosity. The matter ejected by the second energy injection collides and interacts with the expanding envelope or matter ejected by the first energy injection (dashed lines in Figure \ref{Velocity}; discontinuities around $M_r \simeq 14.5M_\odot$ in each panel correspond to shock). After that the gravitationally bound part of expanding envelope begins to fall back, and this fall-back matter undergoes damped oscillations (Fig. \ref{trajectory}). The trajectory of a fluid element follows that of a free-fall particle as seen from the red bold line. This also indicates that the gravity dominates the motion of the fall-back matter. Different periods of oscillation between each radius cause collisions (e.g., collision between $M_r = 12.5M_\odot$ and $12.0M_\odot$ at $t = 2.3 \times 10^{6}\ \mathrm{s}$ in Fig. \ref{trajectory}). These kinds of collisions result in additional shock waves (e.g., solid line at $M_r = 12.3M_\odot$ and $13.4M_\odot$ in bottom panel of Fig. \ref{Velocity}) and discontinuity in density profiles (solid lines in each panel of Fig. \ref{Density}).

There are three key differences between the first and second mass eruptions. First, the brightening phase associated with the second mass eruption lasts more than ten times longer than that in the first mass eruption (Fig. \ref{RSG_LC} and \ref{BSG_LC}). The luminosity gradually increases because photons can easily diffuse out from the shock wave in the inflated low-density envelope. After the peak, the luminosity slowly declines and the brightening phase lasts $\sim 100$ days in model RSG1-f,s,m,l,2-f and $\sim$ ten days in model BSG1-f,s,m,l,2-f. We describe what makes the different duration of brightening phases between two eruptions in section 3.3.

Second, there is a significant difference in the color between the first and second mass eruption events (Fig. \ref{RSG_TEMP} and \ref{BSG_TEMP}). For every model, when the first mass eruption takes place, its effective temperature rises by a factor of $\sim$ ten. On the other hand, in the second mass eruption, the effective temperature rises but does not reach 10,000 K. The temperature at the local maximum in the second mass eruption is even lower than the effective temperature of the original state before the first energy injection (namely, the output value from MESA). Thus a progenitor star observed a few years before the SN does not necessarily follow the standard core mass luminosity relation constructed by the standard stellar evolution models that assume quasi-hydrostatic evolution.

Third, the amount of ejected mass is different between the first and second mass eruption even with the same injected energy ($E_\mathrm{inj1} = E_\mathrm{inj2}$ for RSG1-f,s,m,l and BSG1-f,s,m,l). Interestingly, RSG1-f and BSG1-f show the opposite results. While the second energy injection ejects $\sim 2.5$ times higher mass than the first injection in model BSG1-f, it ejects only one-tenth mass of the first injection in model RSG1-f. These results come from two antagonism effects, which are the weakly bound envelope after the first eruption (this effect enhances eruption) and shock wave attenuation due to diffusion of photons (this effect weakens eruption). While the latter effect works in model RSG1-f, the former effect works more effectively in model BSG1-f.

The amount of ejected mass is sensitive to energy input. In model RSG2-f, the second energy injection ejects $\sim 5 \times 10^2$ times higher mass than that in model RSG1-f,s,m,l even though the difference of $E_\mathrm{inj2}$ between them is only a factor of two. This is because the velocity of matter where the shock wave locates ($v_\mathrm{matter}$) and the escape velocity ($v_\mathrm{esc}$) are very close to each other in cases in which a part of the envelope is ejected. A certain part of the envelope is marginally bound ($v_\mathrm{matter} \simeq v_\mathrm{esc}$) and a slight difference in energy input and resulting shock velocity determine whether it is bound or unbound. For example, when we compare second eruptions in RSG1-f and RSG2-f at $t =  2.5 \times 10^{7}\ \mathrm{s}$, $v_\mathrm{matter} \simeq 4 \times 10^{6}\ \mathrm{cm\ s^{-1}}$ for RSG1-f and $v_\mathrm{matter} \simeq 6 \times 10^{6}\ \mathrm{cm\ s^{-1}}$ for RSG2-f (dashed lines at top and second top panel in Fig. \ref{Velocity}). On the other hand $v_\mathrm{escape}$ is around $5-6\times 10^{6}\ \mathrm{cm\ s^{-1}}$ in this region. Thus, the magnitude relations between $v_\mathrm{matter}$ and $v_\mathrm{esc}$ are different between RSG1-f and RSG2-f and it makes a large difference in ejected mass (Table \ref{table:3}). This kind of sensitivity implies that, although there are a wide diversity of observed SNe IIn or II-P in terms of the presence or the amount of CSM, this diversity may be caused by a smaller difference in the late-phase progenitor evolution and extra energy supply between different progenitors than has previously been thought.

We have seen the three key differences between the first and second mass eruptions. These differences hold at least for an interval in the range $0.5t_\mathrm{dyn} < \Delta t_\mathrm{inj} < 4.0 t_\mathrm{dyn}$ and would hold over the wider range of $\Delta t_\mathrm{inj}$ because the envelope contracts on the Kelvin-Helmholtz timescale ($\sim 200\ \mathrm{yr}$ for model RSG and BSG), which is much longer than the dynamical timescale. Meanwhile, when we compare the properties of the second eruptions between models that have different intervals of energy injection $\Delta t_\mathrm{inj}$ (-f,-s,-m and -l), a model with a longer interval tends to have a higher peak luminosity (Fig. \ref{intervalCompare}). Correspondingly, the total radiated energy also has the same trend. On the other hand, models that have different intervals eject a similar amount of matter with a similar kinetic energy (Table \ref{table:3}). These behaviors originate from the velocity of decelerated matter once ejected in the first eruption. As the interval becomes longer, the ejected matter in the first eruption is gradually decelerated by the gravity of the star, while matter velocity in the downstream region of the outward-shock front is almost constant because the same amount of energy is injected. Thus, the difference in velocities between the upstream and downstream regions increases for longer intervals (e.g., for models RSG1-s,f,m,l, $1.9\times 10^{6}$, $2.7\times 10^{6}$, $4.3\times 10^{6}$, and $7.9\times 10^{6} \mathrm{\ km\ s^{-1}}$, respectively, at the luminosity peak), leads to higher pressure and temperature in the shocked region, and results in a higher peak luminosity and radiated energy. The  peak luminosity and radiated energy in Fig.\ref{intervalCompare} seem to increase more gradually with increasing $\Delta t_\mathrm{inj}$ and would converge when $\Delta t_\mathrm{inj}$ is large enough that the deceleration and resultant fallback of matter once ejected in the first eruption event settles in.

Each ejected fluid element that has a positive total energy (summation of thermal, gravitational, and kinetic energy) keeps expanding at an almost constant velocity until the core collapse of the star. Erupted material reaches $\sim 2\times 10^{15}\ \mathrm{cm}$ (RSG) or $\sim 3\times 10^{16}\ \mathrm{cm}$ (BSG) from the progenitor when the star undergoes core collapse. Their velocities are several $10^{6}\ \mathrm{cm\ s^{-1}}$ (RSG) or several $10^7\ \mathrm{cm\ s^{-1}}$ (BSG) (Fig. \ref{DensityVelocity}). These values do not contradict observations \citep[e.g., ][]{2012ApJ...744...10K}. When we express the velocity-density relation in Figure \ref{DensityVelocity} as $\rho \propto v^{-s}$, the CSM formed by the first eruption (broken line) have larger $s$ values compared with that by the second ejection (solid line), which roughly follow a line with $s\sim1.5$. This is because a part of photons diffusing out from the shock wave formed by the second ejection can be absorbed and accelerate the inner part of the CSM formed by the first eruption. This "steepening" of $s$ value from $\sim1.5$ (and resulting CSM) can affect the light curve of SNe powered by CSM interaction modeled by \citet{1982ApJ...258..790C}.

\subsection{Compare light curves between the first and second eruptions}

The light curves of first and second eruptions show different features (Fig. \ref{RSG_LC} and \ref{BSG_LC}). In this section, we discuss the difference between the first and second eruptions using model RSG1-f as an example.

In the first eruption, the outward shock wave transports energy without leaking photons until it approaches the surface of the star, and thus we can observe strong shock breakout (Fig. \ref{RSG_LC}). In the second eruption, photons diffuse out from the shock front long before it reaches the photosphere, where $\tau = 1,$ because the first eruption lowers the density from the original (MESA output) state. Thus photons travel longer distances to the photosphere, which leads to a broader peak in the light curve. We can see these diffusing out photons in Fig. \ref{Eject_Detail}. When $t = 2.20 \times 10^{7}\ \mathrm{s}$, the shock wave is located at $M_r = 14.3M_\odot$, but the pre-shocked region ($M_r > 14.3M_\odot$) has already heated compared to $t = 1.85 \times 10^{7}\ \mathrm{s}$ owing to leaking photons from the shock. Correspondingly, the luminosity reaches peak at $t = 2.20 \times 10^{7}\ \mathrm{s}$ (Fig. \ref{RSG_LC}).

The duration of the brightening phase is determined by the photon diffusion timescale. Therefore, firstly we consider the region whose radiation energy can contribute to the light curves and then estimate the diffusion timescale of the region. When the diffusion velocity $v_\mathrm{diffusion}$ in a certain region becomes greater than the wave velocity $v_\mathrm{wave}$, photons can diffuse out from the wave. Since $v_\mathrm{diffusion}$ is written as
\begin{equation}
v_\mathrm{diffusion} \simeq c/(3\tau ),
\end{equation}
the radius $R_\tau$, where $v_\mathrm{diffusion} \simeq v_\mathrm{wave}$ holds at the time of shock breakout, has the optical depth
\begin{equation}
\tau \simeq \frac{c}{3v_\mathrm{wave}}.
\end{equation}
The radiation energy density at the place where $\tau < c/(3v_\mathrm{wave})$ is transferred at the diffusion velocity and contributes the tail part of the light curve. Thus, we can obtain a rough estimate of the emergent radiation flux during the light curve peak. As a result, the luminosity $L$ with the duration $\Delta t$ is estimated as
\begin{eqnarray}
L &=& 4\pi R_{\tau}^2 aT_\mathrm{\tau}^4 \frac{c}{3\tau} = 4\pi R_{\tau}^2 aT_{\tau}^4 v_\mathrm{wave},\\
\Delta t &=& \frac{R_\mathrm{ph} - R_{\tau}}{v_\mathrm{diff}} = \frac{R_\mathrm{ph} - R_{\tau}}{c/(3\tau)}.
\end{eqnarray}
 We obtain $v_\mathrm{wave}\simeq53.4$ km s$^{-1}$, $R_{\tau} \simeq 4.91 \times 10^{13}\ \mathrm{cm}$ and $R_\mathrm{ph} \simeq 4.96 \times 10^{13}\ \mathrm{cm}$ in the first eruption from the results of model RSG1-f. The above equations give $\tau \simeq 1870, L \simeq 4.1 \times 10^{38}\ \mathrm{erg\ s^{-1}}$, and $\Delta t \simeq 1\ \mathrm{days}$. In the second eruption, $v_\mathrm{wave}\simeq53.5$ km s$^{-1}$, $R_{\tau} \simeq 8.43 \times 10^{13}\ \mathrm{cm}$, and $R_\mathrm{ph}\simeq 1.26 \times 10^{14}\ \mathrm{cm}$. Thus, $\tau \simeq 1870, L \simeq 3.2 \times 10^{38}\ \mathrm{erg\ s^{-1}}$, and $\Delta t \simeq 89\ \mathrm{days}$. A large difference in $R_\mathrm{ph} - R_{\tau}$ between the two eruptions causes different behaviors of the two brightening events. After the first eruption, the expansion of the envelope significantly increases $R_\mathrm{ph}$. Thus, photons need to travel longer distances from the shock front to the photosphere. This results in a broader peak of the light curve owing to the second eruption. The mass coordinate, where $\tau \simeq 1870,$ at shock breakout is $M_r \sim 14.6M_\odot$ for the first eruption and $\sim 14.3M_\odot$ for the second eruption (Fig. \ref{Eject_Detail}), and thus, the enclosed mass range that contributes to the light curve is also wider in the second eruption compared with the first eruption.

\begin{figure}
\resizebox{\hsize}{!}{\includegraphics{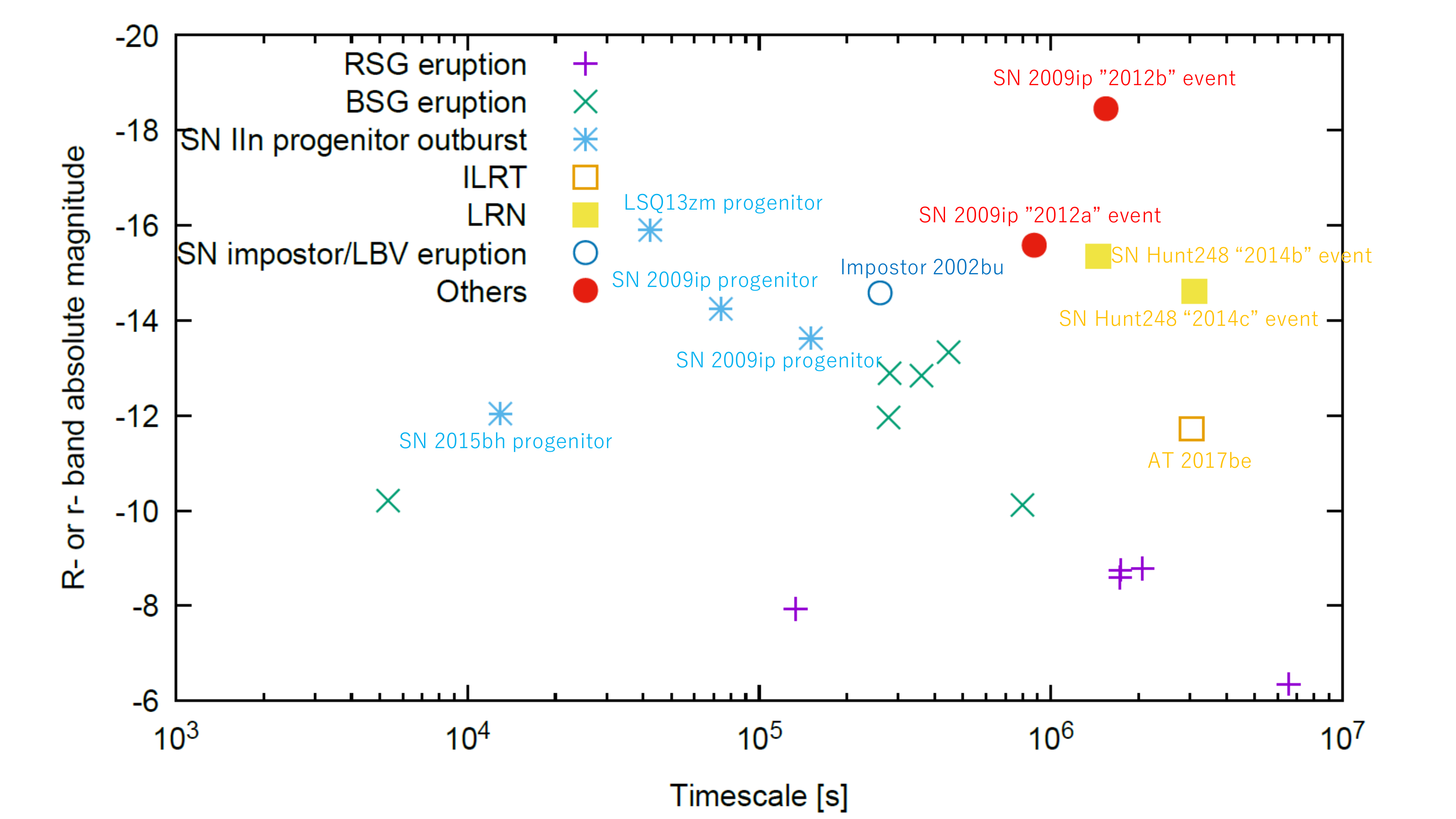}}
\caption{Peak luminosity in R band as a function of the half-decay time of the luminosity for each eruption in our simulation. The bolometric luminosity in our simulation is filtered by Johnson-Cousins R band under the assumption of blackbody radiation. Filter transparency data are taken from the website of Leibniz Institute for Astrophysics Potsdam (AIP)\protect\footnotemark. $3064\ \mathrm{Jy}$ is adopted as the zero magnitude flux in R band  \citep{1998A&A...333..231B}. Some  observational data of SN progenitor outbursts (SN 2009ip \citep{2013ApJ...767....1P, 2013MNRAS.433.1312F}, LSQ13zm \citep{2016MNRAS.459.1039T}, and SN 2015zh \citep{2016ApJ...824....6O}) and gap transients (intermediate luminosity red transient (ILRT) AT 2017be \citep{2018MNRAS.480.3424C}, luminous red nova (LRN) SN Hunt248 \citep{2015MNRAS.447.1922M,2015A&A...581L...4K}, and SN impostor SN 2002bu \citep{2004PASJ...56S...1K}) in R band or r band are also plotted for comparison. Half-decay times of the luminosities for these transients are estimated by linear interpolation of observed magnitudes as a function of time. SN 2009ip shows a lot of peaks (events) in the light curve during 2009-2012. Four representative and well-observed pre-SN outbursts at MJD$\sim$ 55098, 55826, and the two events (2012a and 2012b) introduced in Section 1, are chosen. SN Hunt248, which is thought to be a merger-burst event \citep{2018MNRAS.473.3765M} and classified as LRN \citep[e.g.,][]{2017MNRAS.471.3200M, 2019A&A...630A..75P}, shows triple peaks (outburst events) referred to as 2014a, b, and c \citep{2015A&A...581L...4K}. The 2014b event had begun before the luminosity of the 2014a event declined by a factor of 2 from the peak. Accordingly, only the 2014b and c events are plotted. It should be noted that objects in this figure are representative and that this figure does not show all of the known pre-SN outbursts or gap transients.}
\label{CompareObservation}
\end{figure}
\footnotetext{https://www.aip.de/en/research/facilities/stella/instruments/data}

\section{Discussion}
Our results show unambiguous differences between the first and second mass eruption in terms of light curves, colors, and amounts of erupted unbound mass. Therefore, when we discuss the observation of pre-CCSN mass eruption events and compare them with some physical models or simulations, we may have to take into consideration the effect of repeated mass eruptions seen in our simulations and the altered density profile of the envelope. It should be noted that our results might be restricted to stars with an initial mass of $15M_\odot$. It is important to explore multiple eruption events from stars with different masses, although this is beyond the scope of the present paper.

The motivation of this work is to interpret the nature of multiple mass eruption events accompanied by rapid and repeated luminosity variance. In case of erratic pre-CCSN phase of SN 2009ip during 2011 March-November, its magnitude oscillated with an amplitude of about three mag \citep{2013ApJ...767....1P}. The period of this luminosity oscillation is less than 50 days and this phase lasts about ten months. The peak magnitude and the time interval between each local peak are almost unchanged during this phase. Although the progenitor mass of SN 2009ip ($50-80M_\odot$ \citep{2010AJ....139.1451S}) and our models ($15M_\odot$) are quite different, these observed properties seem to be incompatible with our results. In our work, once the first mass eruption occurs, the density structure of the envelope is altered. Thus the local peak luminosity and color are significantly different between the first and second mass eruption (Fig. \ref{RSG_LC}, \ref{BSG_LC}, \ref{RSG_TEMP} and \ref{BSG_TEMP}), although the same amount of energy is deposited on the same timescale (model RSG1-f,s,m,l and BSG1-f,s,m,l). In addition, our models predict a slower decline of the luminosity after the second local peak because of the extended envelope as shown in Figure \ref{CompareObservation}, where the peak luminosity of each eruption in R band is plotted as a function of half-decay time of the luminosity and some observations are plotted for comparison. A typical half-decay time of the luminosity in models BSG is $\sim 5 \times 10^{3} \mathrm{\ s}$ for the first eruption but $\sim$ several $10^{5} \mathrm{\ s}$ for the second eruption, which is longer than that of SN 2009ip progenitor outbursts in 2009 and 2011.
If these behaviors in our model hold for a more massive star including the  progenitor of SN 2009ip (though it is not obvious), our results may suggest that the nature of mass eruption from SN 2009ip in 2011 cannot be explained by the spherically symmetric eruption of a single star triggered by extra energy supply into the envelope.

Mass eruptions treated in this paper show a wide variety of timescale and peak luminosity depending on the progenitor type (RSG or BSG) and "history of eruptions" (first or second eruption) (Fig. \ref{CompareObservation}). They tend to be a bit darker than the known "gap transients" \citep{2012PASA...29..482K} but brighter than classical novae. If $\sim 15M_\odot$ stars experience these kinds of eruptions prior to core collapse, upcoming deep and high cadence transient surveys will fill in the new gap region between the known gap transients and classical novae.

Observations have revealed the diversity of SNe IIn \citep{2013A&A...555A..10T, 2017hsn..book..403S} in the amount, composition, position, and morphology of CSM. In particular, there is a wide range of variety in the amount of CSM. While some progenitors are associated with more than $30-50M_\odot$ (SN 2010jl; \citet{2012AJ....144..131Z}) or $20 M_\odot$ (SN 2006gy; \citet{2010ApJ...709..856S}) of CSM, others only have $0.003 M_\odot$ (SN 1998S; \citet{2001MNRAS.325..907F}) or less than $0.1 M_\odot$ (SN 2005gl; \citet{2009Natur.458..865G}). This variety could originate from different mechanisms of eruption to form CSM, of course, from a different progenitor mass. As introduced in Sect. 1, there are a wide variety of scenarios that may explain extra energy supply and resulting mass loss and CSM formation. On the other hand, our results show that only a factor of two difference in the amount of deposited energy results in a difference of ejected mass by more than two orders of magnitude (the second mass eruption in RSG1-f and RSG2-f, Table \ref{table:3}). Thus a wide range of the amount of CSM might also reflect a small difference in the amount of extra energy supplied.

When we compare the results of models RSG and BSG, we find that model BSG requires more than ten times larger amount of energy than model RSG to eject the same mass; both  models RSG2-f and BSG1-f expel $\sim 0.7M_\odot$ throughout double eruption events. From this aspect, model RSG (i.e., stars in a relatively metal-rich environment) is preferred as a progenitor of SNe IIn if we only focus on the envelope density stratification and ignore the metallicity dependence of the physics involved in the extra energy injection. At a glance, our results seem to be consistent with \citet{2019arXiv190513197G}, who suggests that SNe IIn prefer relatively higher metallicity environment compared with the other subtypes of SNe II based on a database analysis work. However, there should be various other physical factors correlated with metallicity and affecting the classification of a (sub-)type of an SN. Therefore, our result that metal-rich stars seem to easily suffer from mass eruption is just one of the possible factors SNe IIn prefer.

\section{Conclusions}
We have carried out radiation hydrodynamical simulations to investigate the properties of the repeated dynamical mass eruption events from a single $15M_\odot$ star prior to the core collapse, in the context of the formation of  CSM around SNe IIn. The key point of this work is that we caused mass eruption events not only once but twice by depositing energy two times. The dynamical properties of single mass eruption event have already been studied by \citet{2020A&A...635A.127K}, \citet{ 2019MNRAS.485..988O}, and \citet{ 2010MNRAS.405.2113D}. However, recent observations of SN IIn progenitors indicate that mass eruption phase often occurs more than once just before core collapse. A mass eruption event can alter the density structure of the envelope and therefore the subsequent mass eruption event can show completely different observational features. From this view point, we deposited the extra energy into the bottom of the progenitor envelope twice and simulated the time evolution of both eruption events. The progenitor models were made by MESA. The amount of deposited energy is chosen to reproduce observed CSM, and roughly set to one-third or one-quarter of $|E_\mathrm{envelope}|$ (binding energy of the envelope). We did not deal with the origin of extra energy, which has not been completely understood so far, although it should be studied in our future work.

Results show non-negligible differences between the first and second mass eruption events. After the first eruption event occurs, the envelope is temporarily inflated on the Kelvin-Helmholtz timescale. Thus, when the second mass eruption takes place consecutively in the inflated envelope, it shows a fainter, redder, and long-term eruption. This result may conflict with observations for SN 2009ip during the outburst phase in 2011, although the progenitor mass and properties of the envelope in SN 2009ip are quite different from those of our models. The symmetric mass eruption model from a single star could be ruled out for this event in SN 2009ip if our claims about the properties of second eruption hold for more massive stars. Erupted material from the second eruption collides with the prior CSM from the first eruption and alter the density profile of the prior CSM as described in Sect.3.2. This interaction can affect the light curve of the subsequent CSM interacting SN. Another suggestion of our work is that only a few factors of difference in extra energy makes a larger (by a few orders of magnitude) difference in the amount of erupted mass. In other words, a little difference in the progenitor stellar evolution can be amplified and emerges as a significant difference in the observational features during the pre-CCSN phase and SN itself. From this view point, the difference of the progenitor star evolution between SNe IIn and normal SNe II could be smaller than we think. The peak luminosities of eruptions from $15M_\odot$ stars in this paper lie in between that of known gap transients and classical novae. Higher cadence and deeper transient survey in the future will provide us a larger amount of detailed pre-SN activity data and some objects will show multi-peaked luminosity fluctuations. Interpreting these would require the inclusion of the effect of multiple eruptions and the corresponding altered density structure discussed in this work.

\begin{acknowledgement}
We thank the anonymous referee for giving us a lot of valuable comments and Amy Mednick for language editing. This work is partially supported by JSPS KAKENHI NOs. 16H06341, 20H05639, MEXT, Japan.
\end{acknowledgement}

\bibliography{citation}

\begin{appendix}
\section{Detailed methods of making the two progenitor models using MESA}
The two progenitor models, RSG and BSG, which were used in our simulation as the initial models, were made via a stellar evolution code MESA release 10398 \citep{2011ApJS..192....3P, 2013ApJS..208....4P, 2015ApJS..220...15P, 2018ApJS..234...34P, 2019ApJS..243...10P}.

According to Table \ref{table:1}, initial mass and metallicity were set to the following values and we started to make these models evolve.
\begin{lstlisting}[basicstyle=\ttfamily\footnotesize, frame=single]
    initial_mass = 15.0d0
    initial_z = 0.02d0 (for model RSG)
    Zbase = 0.02d0 (for model RSG)
    initial_z = 0.0002d0 (for model BSG)
    Zbase = 0.0002d0 (for model RSG)
\end{lstlisting}

After the termination of main sequence, we set the following options which designate the method of opacity and mass loss.
\begin{lstlisting}[basicstyle=\ttfamily\footnotesize, frame=single]
    use_Type2_opacities = .true.
    cool_wind_RGB_scheme = 'Dutch'
    cool_wind_AGB_scheme = 'Dutch'
    RGB_to_AGB_wind_switch = 1d-4
    Dutch_scaling_factor = 0.8
\end{lstlisting}
Parameters other than those above were set to default values throughout the entire calculation.
We stopped the calculation 11.2 yr before core collapse (model RSG) and 7.2 yr before core collapse (model BSG) and adopted as initial models for our calculations.
\end{appendix}

\end{document}